\documentclass[12pt]{article}           
\begin{document}
\begin{center}
Symplectic algorithm for systems with second-class constraints
\end{center}
\begin{center}
{\"{O}zlem Defterli\footnote{E-mail: defterli@cankaya.edu.tr }}\\
{Department of Mathematics and Computer Sciences, Faculty of Arts
and Science, \c{C}ankaya University, Balgat 06530, Ankara,
Turkey}\\
{Dumitru Baleanu\footnote{E-mails: dumitru@cankaya.edu.tr,
baleanu@venus.nipne.ro}}\\
 {Department of Mathematics and Computer
Sciences, Faculty of Arts
and Science, \c{C}ankaya University, Balgat 06530, Ankara, Turkey\\
and\\Institute of Space Sciences, P.O.BOX, MG-23, R 76900,
Magurele-Bucharest, Romania}
\end{center}

\begin{abstract}
The recently modified Faddeev-Jackiw formalism for systems having
 one chain of four levels of only second-class constraints is applied to the non-trivial a=1 bosonized chiral Schwinger model in (1+1)
dimensions as well as to one mechanical system. The sets of
obtained constraints are in agreement with Dirac's canonical
formulation.
\end{abstract}

\section{Introduction}
The first order Lagrangian \cite{Shirzad05} to start with is given
by
\begin{equation}\label{Lagrangian1}
L=c_{\alpha}(\zeta){\dot \zeta^\alpha}- H(\zeta),
\end{equation}
where $\zeta^{\alpha}$, $\alpha=1,\ldots,2N$ are the coordinates
and $H(\zeta)$ is the corresponding canonical Hamiltonian. The
equations of motion can be written as $f_{\alpha\beta}{\dot
\zeta^\beta}=
\partial_\alpha H$, where $\partial_\alpha\equiv\partial/\partial\zeta^\alpha$ and
\begin{equation}\label{tensor1}
f_{\alpha\beta}\equiv \partial_\alpha
c_\beta(\zeta)-\partial_\beta c_\alpha(\zeta).
\end{equation}
The system corresponding to the Lagrangian (\ref{Lagrangian1})
together with the primary constraints $\phi_\mu^{(1)}$,
$(\mu=1,\ldots,M)$, is described by the Lagrangian $L'=
c_{\alpha}{\dot \zeta^\alpha}-\lambda^\mu\phi_\mu^{(1)} -
H(\zeta)$ where $\lambda^\mu$ are Lagrange multipliers. The
consistency conditions of these primary constraints bring new
constraints to the system. So, the Lagrangian (\ref{Lagrangian1})
is extended by the term $\xi^\mu\dot{\phi}^{(1)}_\mu$
 as $L^{(1)}= c_{\alpha}(\zeta){\dot
\zeta^\alpha}-\dot{\xi}^\mu\phi_\mu^{(1)} - H(\zeta)$
\cite{Rothe03}. Then, multiplying both sides of the associated
equations of motion by the left null-eigenvectors of the
symplectic tensor gives the next second-class constraint and so
on. This procedure leads us to a chain structure \cite{Loran02} of
a second-class constraint set $\phi^{(1)}, \phi^{(2)},\ldots,
 \phi^{(N)}$.
Assume that the above method continues until the step $(N/2)+1$
because of the singularity of the symplectic two-form
$F^{((N/2)+1)}$ given in the right hand side of the matrix form of
the following equations of motion \cite{Shirzad05}
\begin{eqnarray}\label{eqnmotion}
  \left( \begin{array}{cccc}
f&A^{(1)}& \ldots &A^{[(N/2)+1]}\\
-A^{(1)T}&0&\ldots &0\\
\vdots&\vdots & &\vdots \\
-A^{[(N/2)+1]T}&0& \ldots &0\\
\end{array}\right)
\left(%
\begin{array}{c}
  {\dot \zeta} \\
  {\dot \xi_1} \\
\vdots \\
   {\dot \xi_{(N/2)+1}}\\
\end{array}%
\right)=\left(%
\begin{array}{c}
   \partial H \\
  0 \\
   \vdots\\
  0 \\
\end{array}%
\right),
\end{eqnarray}
where the elements of the matrix $A^{(\gamma)}$ are $
A^{(\gamma)}_{\mu\alpha}\equiv
\partial_\alpha \phi_\mu^{(\gamma)}$, $\gamma=1, \ldots, N$. So, $F^{((N/2)+1)}$ does not admit a new
left null-eigenvector. In order to obtain the next level
constraint $\phi^{[(N/2)+2]}\approx \{\phi^{[(N/2)+1]}, H \}$ and
the remaining constraints of the chain, a modification in the
symplectic analysis \cite{Neto92b, Neto94} given above is done
\cite{Shirzad05} by truncating the matrix form of the symplectic
tensor $F^{((N/2)+1)}$ in (\ref{eqnmotion}) as
\begin{eqnarray}\label{truncatedF}
 \tilde{F}^{((N/2)+1)}=\left( \begin{array}{cc}
f&A^{(1)} \\
-A^{(1)T}&0\\
\vdots & \vdots\\
-A^{[(N/2)+1]T}&0\\
\end{array}\right)\end{eqnarray}
so that $\tilde{F}^{((N/2)+1)}$ possesses a new left
null-eigenvector $v^{[(N/2)+1]}$. Here, $\tilde{F}^{((N/2)+1)}$
denotes the truncated matrix $F^{((N/2)+1)}$ \cite{Shirzad05}.
\\The aim of this
manuscript is to apply recently modified
Faddeev-Jackiw\cite{Faddeev88} formalism on two examples of
physical interest. The first example is the Schwinger model in
(1+1) dimensions which, despite the gauge anomaly, is unitary and
it was consistently quantized \cite{Kim06}. The second example is
a simpler mechanical system possessing one chain of a four level
second-class constraints.

\section{Examples}
\subsection{ Schwinger model in (1+1) dimensions}
The bosonized chiral Schwinger model in (1+1) dimensions is
described by the Lagrangian density
\begin{eqnarray}\label{example2L}
 L = \frac{1}{2}\partial_\mu\varphi\partial^\mu\varphi + ( g^{\mu\nu}-\varepsilon^{\mu\nu})\partial_\mu\varphi A_\nu -
 \frac{1}{4}F_{\mu\nu}F^{\mu\nu}+ \frac{1}{2}A_\mu A^\mu,
\end{eqnarray}
where $g^{\mu\nu}$=diag(1, -1), $\varepsilon^{01}$=1 and
$\partial_0\equiv\partial/\partial t$,
$\partial_1\equiv\partial/\partial x$. The canonical momenta are
$\pi^0 = 0$, $\pi^1 = \dot{A_1}-\partial_1A_0\equiv E$ and
$\pi_\varphi = \dot{\varphi}+ A_0 - A_1 \equiv \pi$. The canonical
density  and the total density Hamiltonians are given \cite{Kim06}
by $H_C = \frac{1}{2}(E^2 + \pi^2+(\partial_1\varphi)^2)+
E\partial_1 A_0+
 (\pi+A_1+\partial_1\varphi)(A_1 - A_0)$ and $H_T = H_C + \lambda
 \pi^0$, respectively, where the overdot denotes the time derivative. The usual Dirac
approach \cite{dirac} gives the following set of constraints
\cite{Kim06} $\phi^{(1)}\equiv\pi^0\approx0$, $\phi^{(2)}\equiv
\partial_1 E + \pi+\partial_1\varphi+A_1\approx0$, $\phi^{(3)}\equiv E\approx0$, and
$\phi^{(4)}\equiv-\pi-\partial_1\varphi-2A_1+A_0\approx0$ where
$\phi^{(1)}\equiv\pi^0$ is the only primary constraint and all
constraints are second-class. The first order Lagrangian density
associated to (\ref{example2L}) is written as
\begin{eqnarray}\label{example2firstL}
 L= \pi^0\dot{A_0}+\pi^1\dot{A_1}+\pi_{\varphi} \dot{\varphi}- H_C.
\end{eqnarray}
which gives the symplectic tensor $f$ as follows
\begin{eqnarray}\label{example2f}
 f=\left( \begin{array}{cccccc}
0&0&0&-1&0&0\\
0&0&0&0&-1&0\\
0&0&0&0&0&-1\\
1&0&0&0&0&0\\
0&1&0&0&0&0\\
0&0&1&0&0&0\\
\end{array}\right)\delta(x-y). \end{eqnarray}
Considering the consistency condition of the primary constraint
$\phi^{(1)}\equiv\pi^0$ forms the first order Lagrangian density
(\ref{example2firstL}) as
\begin{eqnarray}\label{example2L1}
 L^{(1)} =
 \pi^0\dot{A_0}+\pi^1\dot{A_1}+\pi_\varphi\dot{\varphi}-\dot{\xi_1}\pi^0-H_C
\end{eqnarray} and so by eqn.(\ref{tensor1}) and (\ref{eqnmotion}) the equations of motion $F^{(1)}_{\alpha\beta}{\dot \zeta}^\beta =
\frac{\delta H_T}{\delta\zeta^\alpha}$ in the extended space
$\zeta=(A_0, A_1, \varphi, \pi^0, \pi^1, \pi_\varphi, \xi_1)$ are
written as in the following
\begin{eqnarray}\label{example2-eqnmotion1}
  \left( \begin{array}{ccccccc}
0&0&0&-1&0&0&0\\
0&0&0&0&-1&0&0\\
0&0&0&0&0&-1&0\\
1&0&0&0&0&0&1\\
0&1&0&0&0&0&0\\
0&0&1&0&0&0&0\\
0&0&0&-1&0&0&0\\
\end{array}\right)
\left(%
\begin{array}{c}
  {\dot A_0} \\
  {\dot A_1} \\
  {\dot \varphi} \\
  {\dot \pi^0} \\
  {\dot \pi^1} \\
  {\dot \pi_\varphi} \\
  {\dot \xi_1} \\
\end{array}%
\right)=\left(%
\begin{array}{c}
   -(\pi+A_1+\partial_1\varphi+ \partial_1\pi^1)\\
 \pi+A_1+\partial_1\varphi+A_1-A_0\\
  -\partial_1A_1+\partial_1A_0  \\
  \lambda\\
   \pi^1+\partial_1A_0\\
  \pi_\varphi+A_1-A_0\\
  0 \\
\end{array}%
\right),
\end{eqnarray}
where the left null-eigenvector of $F^{(1)}$ is calculated as
$v^{(1)}=( -1, 0, 0, 0, 0, 0, 1 )$. By applying the procedure with
$v^{(1)}$, the second constraint is found to be $\phi^{(2)}=
\pi+A_1+\partial_1\varphi+ \partial_1\pi^1$ that is second-class.
Then the consistency of $\phi^{(2)}$ adds the term $
-\dot{\xi_2}(\pi+A_1+\partial_1\varphi+ \partial_1\pi^1)$ to the
Lagrangian in (\ref{example2L1}) and from (\ref{eqnmotion}) the
matrix form of the equations of motion $F^{(2)}_{\alpha\beta}{\dot
\zeta}^\beta = \frac{\delta H_T}{\delta\zeta^\alpha}$ is found as
follows
\begin{eqnarray}\label{example2-eqnmotion2}
  \left( \begin{array}{cccccccc}
0&0&0&-1&0&0&0&0\\
0&0&0&0&-1&0&0&1\\
0&0&0&0&0&-1&0&0\\
1&0&0&0&0&0&1&0\\
0&1&0&0&0&0&0&0\\
0&0&1&0&0&0&0&1\\
0&0&0&-1&0&0&0&0\\
0&-1&0&0&0&-1&0&0\\
\end{array}\right)
\left(%
\begin{array}{c}
  {\dot A_0} \\
  {\dot A_1} \\
  {\dot \varphi} \\
  {\dot \pi^0} \\
  {\dot \pi^1} \\
  {\dot \pi_\varphi} \\
  {\dot \xi_1} \\
{\dot \xi_2} \\
\end{array}%
\right)=\left(%
\begin{array}{c}
   -(\pi+A_1+\partial_1\varphi+ \partial_1\pi^1)\\
 \pi+A_1+\partial_1\varphi+A_1-A_0\\
  -\partial_1A_1+\partial_1A_0  \\
  \lambda\\
   \pi^1+\partial_1A_0\\
  \pi_\varphi+A_1-A_0\\
  0 \\
  0\\
\end{array}%
\right),
\end{eqnarray}
 where $F^{(2)}$ admits a new left
null-eigenvector $v^{(2)}=( 0, 0, -1, 0, 1, 0, 0, 1 )$. Repeating
the procedure for $v^{(2)}$ gives the next second-class constraint
$\phi^{(3)}= \pi^1$. For the following step we obtain
\begin{eqnarray}\label{example2L3}
 L^{(3)}=
 \pi^0\dot{A_0}+\pi^1\dot{A_1}+\pi_\varphi\dot{\varphi}-\dot{\xi_1}\pi^0-\dot{\xi_2}(\pi+A_1+\partial_1\varphi+
 \partial_1\pi^1)-\dot{\xi_3}\pi^1-H_C.
\end{eqnarray} The form of the eqn.(\ref{eqnmotion}) for this case becomes
\begin{eqnarray}\label{example2-eqnmotion3}
  \left( \begin{array}{ccccccccc}
0&0&0&-1&0&0&0&0&0\\
0&0&0&0&-1&0&0&1&0\\
0&0&0&0&0&-1&0&0&0\\
1&0&0&0&0&0&1&0&0\\
0&1&0&0&0&0&0&0&1\\
0&0&1&0&0&0&0&1&0\\
0&0&0&-1&0&0&0&0&0\\
0&-1&0&0&0&-1&0&0&0\\
0&0&0&0&-1&0&0&0&0\\
\end{array}\right)
\left(%
\begin{array}{c}
  {\dot A_0} \\
  {\dot A_1} \\
  {\dot \varphi} \\
  {\dot \pi^0} \\
  {\dot \pi^1} \\
  {\dot \pi_\varphi} \\
  {\dot \xi_1} \\
{\dot \xi_2} \\
{\dot \xi_3} \\
\end{array}%
\right)=\left(%
\begin{array}{c}
   -(\pi+A_1+\partial_1\varphi+ \partial_1\pi^1)\\
 \pi+A_1+\partial_1\varphi+A_1-A_0\\
  -\partial_1A_1+\partial_1A_0  \\
  \lambda\\
   \pi^1+\partial_1A_0\\
  \pi_\varphi+A_1-A_0\\
  0 \\
  0\\
  0\\
\end{array}%
\right)
\end{eqnarray}
but $F^{(3)}$ does not have a new left null-eigenvector.
 Finally, by truncating the above mentioned symplectic two-form  as shown in
 (\ref{truncatedF}), we obtain
 \begin{eqnarray}\label{example2truncatedF3}
\tilde{F}^{(3)}= \left( \begin{array}{ccccccc}
0&0&0&-1&0&0&0\\
0&0&0&0&-1&0&0\\
0&0&0&0&0&-1&0\\
1&0&0&0&0&0&1\\
0&1&0&0&0&0&0\\
0&0&1&0&0&0&0\\
0&0&0&-1&0&0&0\\
0&-1&0&0&0&-1&0\\
0&0&0&0&-1&0&0\\
\end{array}\right)\delta(x-y)\end{eqnarray}
which admits a new left null-eigenvector $v^{(3)}=( 0, -1, 0, 0,
0, 0, 0, 0, 1 )$. Therefore, the last step of the algorithm
produces the last constraint
$\phi^{(4)}=-\pi-\partial_1\varphi-2A_1+A_0$. By direct
calculations we see that the next step of the algorithm will not
produce any new constraint because the corresponding symplectic
two-form has a non-singular matrix with determinant 1.
\subsection{Example 2 }
Let us consider the following Lagrangian
\begin{equation}\label{ex1L}
L={\dot x}{\dot y}- z( x + y),
\end{equation}
where $\zeta^{M}=( x, y, z, p_x, p_y, p_z ), M= 1,2,3$. The
primary constraint is $\phi^{(1)}=p_z$. The canonical and total
Hamiltonians are given by $H_C = p_x p_y + z( x + y ), H_T = H_C +
\lambda p_z$. By the usual Dirac approach\cite{dirac}, we obtain
the following second-class constraints $\phi^{(2)}= - x - y$,
$\phi^{(3)}= p_x + p_y$ and $\phi^{(4)}= -2z$. The first order
Lagrangian of the form
 (\ref{Lagrangian1}) is
\begin{eqnarray}\label{example1L}
 L={\dot x}p_x + {\dot y}p_y + {\dot z}p_z - H_C.
\end{eqnarray}
After imposing the consistency condition of $\phi^{(1)}$, the
first order Lagrangian in (\ref{example1L}) becomes
\begin{eqnarray}\label{example1L1}
 L^{(1)}={\dot x}p_x + {\dot y}p_y + {\dot z}p_z - {\dot \xi_1}p_z -
 H_C.
\end{eqnarray}
From (\ref{tensor1}) and (\ref{eqnmotion}), the matrix form of the
equations of motion 
is found as
\begin{eqnarray}\label{example1-eqnmotion1}
  \left( \begin{array}{ccccccc}
0&0&0&-1&0&0&0\\
0&0&0&0&-1&0&0\\
0&0&0&0&0&-1&0\\
1&0&0&0&0&0&0\\
0&1&0&0&0&0&0\\
0&0&1&0&0&0&1\\
0&0&0&0&0&-1&0\\
\end{array}\right)
\left(%
\begin{array}{c}
  {\dot x} \\
  {\dot y} \\
  {\dot z} \\
  {\dot p_x} \\
  {\dot p_y} \\
  {\dot p_z} \\
  {\dot \xi_1} \\
\end{array}%
\right)=\left(%
\begin{array}{c}
   z \\
  z \\
   x+y \\
  p_y \\
   p_x \\
 \lambda \\
  0 \\
\end{array}%
\right).
\end{eqnarray}
The matrix $F^{(1)}$ in (\ref{example1-eqnmotion1}) admits the
left null-eigenvector $v^{(1)}= ( 0, 0, -1, 0, 0, 0, 1 )$.
Applying the method described in Section 1, the second-class
constraint $\phi^{(2)}= -x -y$ appears. Imposing the consistency
condition of $\phi^{(2)}$ we obtain
\begin{eqnarray}\label{example1L2}
 L^{(2)}={\dot x}p_x + {\dot y}p_y + {\dot z}p_z - {\dot \xi_1}p_z - {\dot \xi_2}(-x-y)-
 H_C
\end{eqnarray}
so that by using (\ref{eqnmotion}) the corresponding equations of
motion are obtained in the following matrix form
\begin{eqnarray}\label{example1-eqnmotion2}
  \left( \begin{array}{cccccccc}
0&0&0&-1&0&0&0&-1\\
0&0&0&0&-1&0&0&-1\\
0&0&0&0&0&-1&0&0\\
1&0&0&0&0&0&0&0\\
0&1&0&0&0&0&0&0\\
0&0&1&0&0&0&1&0\\
0&0&0&0&0&-1&0&0\\
1&1&0&0&0&0&0&0\\
\end{array}\right)
\left(%
\begin{array}{c}
  {\dot x} \\
  {\dot y} \\
  {\dot z} \\
  {\dot p_x} \\
  {\dot p_y} \\
  {\dot p_z} \\
  {\dot \xi_1} \\
 {\dot \xi_2} \\

\end{array}%
\right) = \left(%
\begin{array}{c}
   z \\
  z \\
   x+y \\
  p_y \\
   p_x \\
 \lambda \\
  0 \\
  0\\
\end{array}%
\right).
\end{eqnarray}
In this case $F^{(2)}$ admits a new left null-eigenvector
$v^{(2)}= ( 0, 0, 0, -1, -1, 0, 0, 1 )$. The same procedure with
$v^{(2)}$ gives the second-class constraint $\phi^{(3)}= p_y +
p_x$. In order to obtain the next constraint $\phi^{(4)}$, the
consistency condition of the constraint $\phi^{(3)}$  is
considered, so the Lagrangian (\ref{example1L2}) takes the form
\begin{eqnarray}\label{example1L3}
 L^{(3)}={\dot x}p_x + {\dot y}p_y + {\dot z}p_z - {\dot \xi_1}p_z - {\dot \xi_2}(-x-y)- {\dot \xi_3}(p_y + p_x)-
 H_C
\end{eqnarray}
which gives
\begin{eqnarray}\label{example1F3}
 F^{(3)}=\left( \begin{array}{ccccccccc}
0&0&0&-1&0&0&0&-1&0\\
0&0&0&0&-1&0&0&-1&0\\
0&0&0&0&0&-1&0&0&0\\
1&0&0&0&0&0&0&0&1\\
0&1&0&0&0&0&0&0&1\\
0&0&1&0&0&0&1&0&0\\
0&0&0&0&0&-1&0&0&0\\
1&1&0&0&0&0&0&0&0\\
0&0&0&-1&-1&0&0&0&0\\
\end{array}\right). \end{eqnarray}
 $F^{(3)}$ does not admit a new left null-eigenvector, therefore a
problem appears to reach the last constraint $\phi^{(4)}$ within
this procedure. Therefore, using (\ref{truncatedF}) we truncate
the matrix $F^{(3)}$ as
\begin{eqnarray}\label{example1truncatedF3}
 \tilde{F}^{(3)}= \left( \begin{array}{ccccccc}
0&0&0&-1&0&0&0\\
0&0&0&0&-1&0&0\\
0&0&0&0&0&-1&0\\
1&0&0&0&0&0&0\\
0&1&0&0&0&0&0\\
0&0&1&0&0&0&1\\
0&0&0&0&0&-1&0\\
1&1&0&0&0&0&0\\
0&0&0&-1&-1&0&0\\
\end{array}\right).\end{eqnarray} $\tilde{F}^{(3)}$
admits a new left null-eigenvector denoted by $ v^{(3)}=( -1, -1,
0, 0, 0, 0, 0, 0, 1 )$. Hence, we found the last constraint
$\phi^{(4)}= -2z$ by continuing the procedure with the
null-eigenvector $v^{(3)}$. The constraint $\phi^{(4)}$ is the
last constraint and the procedure is finished because for the next
level, the symplectic tensor $F^{(4)}$ has a non-singular matrix
with det($F^{(4)}$)=16.

{\small The authors would like to thank the organizers of this
colloquium for giving them the opportunity to attend this meeting.
This paper is partially supported  by the Scientific and Technical
Research Council of Turkey.}


\begin{thebibliography}{99}
\bibitem{Shirzad05} A. Shirzad, M. Mojiri: J. Math. Phys. {\bf46} (2005) 012702.
\bibitem{Rothe03} H. J. Rothe, K. D. Rothe: J. Phys. A {\bf36} (2003) 1671.
\bibitem{Loran02} F. Loran, A. Shirzad: Int. J. Mod. Phys. A {\bf17} (2002) 625.
\bibitem{Neto92b} J. Barcelos-Neto, C. Wotzasek: Int. J. Mod. Phys. A {\bf7} (1992) 4981.
\bibitem{Neto94} J. Barcelos-Neto, N. R. F. Beraga: J. Math. Phys. {\bf35} (1994) 3497.
\bibitem{Faddeev88}
 L. D. Faddeev, R.
Jackiw: Phys. Rev. Lett. {\bf 60}(1988) 1692.
\bibitem{dirac} P. A.
M. Dirac: {\it Lectures on Quantum Mechanics}. Yeshiva University,
New York, 1964.
\bibitem{Kim06}
 Y.-W. Kim, Ee C.-Y., S.-K. Kim, Y.-J. Park: Phys. Lett. B {\bf 632}(2006) 427.
\end{thebibliography}
\end {document}